\documentclass[conference]{IEEEtran}

\makeatletter
\def\ps@headings{%
\def\@oddhead{\mbox{}\scriptsize\rightmark \hfil \thepage}%
\def\@evenhead{\scriptsize\thepage \hfil \leftmark\mbox{}}%
\def\@oddfoot{}%
\def\@evenfoot{}}
\makeatother
\usepackage{hyperref, url} 

\usepackage{amsmath, amssymb, amsfonts, amscd} 

\usepackage{lipsum} 

\usepackage{courier} 

\usepackage{graphicx}
\usepackage{bmpsize}

\usepackage{multirow}

\usepackage{wrapfig}
\usepackage{threeparttable} 
\usepackage{rotating} 
\usepackage{subcaption}
\usepackage{float}
\usepackage{multicol}
\usepackage{colortbl} 
\definecolor{lightred}{rgb}{ .98, .79,  .79} 
\definecolor{darkolivegreen}{rgb}{0.33, 0.42, 0.18} 

\usepackage{color} 
\usepackage[dvipsnames]{xcolor} 

\usepackage{courier} 
\usepackage[pass]{geometry}

\usepackage{pdflscape} 
\usepackage{afterpage}
\usepackage{capt-of}
\usepackage{booktabs} 
\usepackage{arydshln} 

\usepackage{ulem}

\IEEEoverridecommandlockouts 

\begin{document}

\title{
\huge{Towards Malware Detection via CPU Power Consumption:\\Data Collection Design and Analytics (Extended Version)}
\thanks{This is an extended version of the publication appearing in IEEE TrustCom-18 with the same title, including more thorough descriptions of the data collection process and algorithmic details. Please cite the published version.}
\thanks{J.M. Hern\'{a}ndez Jim\'{e}nez led authorship and research involving related work, testbed design and implementation, rootkit analysis, and data collection.}
\thanks{Katerina Goseva-Popstojanova’s work was funded in part by the NSF grant CNS-1618629.}
\thanks{This manuscript has been authored by UT-Battelle, LLC under Contract No. DE-AC05-00OR22725 with the US DOE. The United States Government retains and the publisher, by accepting the article for publication, acknowledges that the United States Government retains a non-exclusive, paid-up, irrevocable, world-wide license to publish or reproduce the published form of this manuscript, or allow others to do so, for United States Government purposes. The DOE will provide public access to these results of federally sponsored research in accordance with the DOE Public Access Plan (\url{http://energy.gov/downloads/doe-public-access-plan)}.}
}


\author{
\IEEEauthorblockN{
Robert Bridges\IEEEauthorrefmark{1},
Jarilyn Hern\'{a}ndez Jim\'{e}nez\IEEEauthorrefmark{1}\IEEEauthorrefmark{2},
Jeffrey Nichols\IEEEauthorrefmark{1},
Katerina Goseva-Popstojanova\IEEEauthorrefmark{2},
Stacy Prowell\IEEEauthorrefmark{1}}
\IEEEauthorblockA{
\IEEEauthorrefmark{1} Computational Science and Engineering Division, Oak Ridge National Laboratory, Oak Ridge, TN 37831 \\
 \{bridgesra, hernandezjm1, nicholsja2, prowellsj\}@ornl.gov}
\IEEEauthorblockA{
\IEEEauthorrefmark{2}Lane Department of Computer Science and Electrical Engineering,
 West Virginia University, Morgantown, WV 26506\\
 \{jhernan7, katerina.goseva\}@mail.wvu.edu}
 }

\maketitle

\thispagestyle{plain}
\pagestyle{plain}
\begin{abstract}
This paper presents an experimental design and data analytics approach aimed at power-based malware detection on general-purpose computers. 
Leveraging the fact that malware executions must consume power, we explore the postulate that malware can be accurately detected via power data analytics.  
Our experimental design and implementation allow for programmatic collection of CPU power profiles for fixed tasks during uninfected and infected states using five different rootkits.  
To characterize the power consumption profiles, we use both simple statistical and novel, sophisticated features. 
We test a one-class anomaly detection ensemble (that baselines non-infected power profiles) and several kernel-based SVM classifiers (that train on both uninfected and infected profiles) in detecting previously unseen malware and clean profiles.  
The anomaly detection system exhibits perfect detection when using all features and tasks, with smaller false detection rate than the supervised classifiers. 
The primary contribution is the proof of concept that baselining power of fixed tasks can provide accurate detection of rootkits. 
Moreover, our treatment presents engineering hurdles needed for  experimentation and allows analysis of each statistical feature individually. 
This work appears to be the first step towards a viable power-based detection capability for general-purpose computers, and presents next steps toward this goal. 
\end{abstract}

\section{Introduction}
\label{sec:intro}
The current ability to protect networked assets from  malicious software (malware) is proving vastly insufficient, which poses a serious national threat as attacks may result in halting critical infrastructure, disclosing state secrets, and financial losses in the billions of dollars.
Signature-based detection malware methods, in which a specific pattern is identified and used to detect attacks with the same pattern, exhibit low false detection rates and are generally the first line of defense.  Yet, they are ineffective against unseen attack patterns and are simply unable to keep pace with the rate and sophistication of modern malware. 
For example, machine-generated malware variants create novel malware in extremely high volume that bypass detection, 
and polymorphic malware avoids detection by regularly rewriting portions of itself so that it is syntactically different but semantically identical.  
In addition to the problem of novel and polymorphic malware side-stepping signature-based methods, a host of other deficiencies exist. 
For instance, signature extraction and distribution is a complex and time consuming task; signature generation involves manual intervention and requires strict code analysis; the size of signature repositories is growing at an alarming rate; signature updates are expensive to implement in terms of downtime and system resources; and most detection software requires highly privileged status to operate, which induces vulnerabilities. 

Current research, typically based on behavior analysis, seeks complementary methods to assist in malware detection. 
A common weakness of these detection methods stems from the fact that they operate on the machine being monitored, which can and has allowed attackers to disable the monitoring software or modify it to prevent detection after gaining entry to the computer. 
This is particularly relevant for rootkits, a type of computer malware that attains administrative privileges, thereby concealing themselves from many detection methods. 
Another line of detection research (including this work) has gained traction in the embedded system space (e.g., programmable logic controllers (PLCs), pharmaceutical compounders, or electric grid synchrophasors) by using out-of-band monitoring of side-channel data (e.g., vibrations, timing, temperature, power, etc.)~\cite{9, gonzalez2009power, gonzalez2011,  gonzalez2014detecting, reed2012, zhong2015side}. 
There is a theoretically sound, fundamental observation upon which these works rely\textemdash if a device's operations are changed by the malicious actor, then some physical consequence must be realized; 
hence, the research question addressed by all such efforts is, ``Can unwanted software be accurately identified by monitoring some physical observable?''. 
Such efforts entail engineering physical sensors and data science techniques to extract the malicious signal from ambient noise. 
Discussed more in the the related work Sec.~(\ref{sec:RelWork}), these advancements generally leverage an element of invariability in the device being protected, e.g., regularity of synchrophasors' signal timing, to limit noise. 
Consequently, their applicability to devices accommodating more variable use (e.g., smart phones, general-purpose computers) is much more difficult (e.g., see Hoffman et al.~\cite{8}) and less explored. 

In this work, we take the first steps to extend power data analysis for malware detection on general-purpose computers. 
Our goal in this work is to prove the concept that malware actions leave a detectable signal in the CPU's power data, and to experiment with statistical learning techniques to measure their efficacy in detection.  
Moreover, we target rootkits\textemdash malware that takes steps to hide its presence,  (e.g., by augmenting the user-visible process list)\textemdash pursuing the hypothesis that masking efforts must alter the power profile, allowing a new avenue for detection.  

To this end, we design an experimental setup consisting of a general-purpose computer running Windows operating system (OS) with out-of-band power monitoring hardware and a programmatic workflow for running a set of tasks both with and without malware present. 
This allows us to monitor and, in a repeatable manner, collect the direct current (DC) power supplying the CPU while performing common tasks under both uninfected and infected states. 
Descriptions of our testbed, experimental design, and engineering challenges incurred comprise Sec.~\ref{sec:expDesignDataColl}.
Initial visual analysis of the power profile of a general-purpose computer, even while sitting idle, shows that the seemingly random or unexpected spikes in power do occur. 
These can be traced back to normal background services, but they illustrate the challenge of our hypothesis under investigation\textemdash 
do the executions of malware during a fixed set of actions produce strong enough changes to the power consumption for accurate detection, even in the presence of the ambient but somewhat unpredictable noise? 
To test this hypothesis, we proceed in Sec.~\ref{sec:analysis} to build a detection capability from statistical data analysis and to test data-analytic techniques to see if the malware is accurately identifiable.

We first propose an unsupervised learning approach that treats this as a one-class problem.  
During training, the algorithm only sees uninfected power profiles that are collected while the machine does a fixed set of tasks. 
The trained detector is then tested on power profiles collected while the same tasks are executed during non-infected and infected states. 
We note that a common problem for anomaly detection (and specifically intrusion detection applications) is the base rate fallacy; e.g., see Axelsson~\cite{axelsson2000base}.\footnote{The base rate fallacy is the presence of both a low false positive rate (percentage of negatives that are misclassified) and a high false detection rate (percentage of alerts that are false positives). 
The base rate fallacy is often caused by high class imbalance (orders of magnitude more negatives than positives), and hence is a chronic issue for anomaly \& intrusion detection. } 
The proposed unsupervised detection method uses an ensemble of single-feature anomaly detectors with the goal of battling the base rate fallacy; specifically, we seek an anomaly detection method that decreases the false detection rate without sacrificing recall (the true-positive rate). 

We also present supervised detection algorithms (several kernel-based support vector machines (SVMs)) equipped with the same features as the anomaly detector. 
For testing we use hold-one-out cross validation. 
That is, the learning algorithm is privy to both uninfected and infected power profiles during training and testing, but the training and testing malware is disjoint. 
Hence, this tests if the supervised detection system can detect never-before-seen malware given observations of both non-infected and infected power profiles.

\subsection{Contributions}
This is the first research work to attempt to detect malware on a general-purpose computer by analyzing the power consumption of the CPU. 
Our testbed and experimental setup, consisting of a general-purpose computer, power-monitoring hardware, and custom software, 
is designed to collect power profiles programmatically with the computer in an uninfected state and while infected by five different rootkits. 
The design, challenges and unique solutions to the data collection framework are research contributions. 
Our anomaly detection ensemble and the results contribute to the anomaly detection research in a variety of ways. 
First, our analysis allows evaluation of each statistical feature's efficacy in malware detection, in particular, showing which features help, hurt, and are inconsequential for accurate detection.  
Second, we provide novel applications of time-series analysis tools, for example, our data-driven approach for transforming time-series power data into a sequence of curve shapes. 
Third, the unsupervised detection results display the quantitative gain in the false detection rate of the ensemble over any of the constituent members with no expense to the true positive rate.  
This shows that the ensemble directly addresses the base rate fallacy and supports a growing body of literature in this area~\cite{bridges2015multi, bridges2016multi, christodorescu2007can, ferragut2011automatic, ferragut2012new, harshaw2016graphprints, moore2017modeling, thomas2010rapid}.
Finally, our anomaly detection ensemble outperforms the supervised algorithms\textemdash perhaps surprising given that the latter is privy to other known malware profiles during training. 

Our results confirm, at least in this experimental setting, that (1) malware does leave a noticeable trace in the power profile, and (2) sophisticated malware can be detected without prior behavior analysis.  
The consequence of our findings is that accurate detection of malware is possible by first baselining a fixed workload and then periodically running that workload to check the new power profile against the baseline. 
Moreover, our rootkit investigations revealed that many otherwise-hard-to-detect rootkits take actions to hide their modifications upon investigations of registries  (e.g,~~\cite{Hoglund08}), process lists, or upon web activity (and presumably caused by other events); hence, this work is a first step to an envisioned technology that baselines the power consumption of a small, fixed set of tasks on an uninfected computer and then uses anomaly detection to identify malware infections.
\section{Related Work}
\label{sec:RelWork-2}
\label{sec:RelWork}
Previous work explored power consumption acquisition aimed at studying power efficiency of servers~\cite{koller2010wattapp, Feng2005} and mobile devices~\cite{Polakis2015PowerslaveAT}. 
Here, we present more details on related works that explored the use of power consumption for malware detection. These works were focused on embedded medical devices~\cite{9}, mobile devices~\cite{8,Azmoodeh2017}, and software-defined radio, PLC and smart grid systems~\cite{gonzalez2009power, gonzalez2011, gonzalez2014detecting, reed2012}. 


Clark et al.~\cite{9} achieved accurate results using supervised classification for malware detection based on power consumption of an embedded medical device and a pharmaceutical compounder (an industrial control workstation). Note that their work is based on monitoring the alternating current (AC) outlet, while we are monitoring direct current (DC) channels. 
The drawback of monitoring AC is a consequence of periodic changes of current direction, which lead to reverses of the voltage, making the analog circuits much more susceptible to noise. 
Also, their approach was tested only on embedded devices, which have a limited set of instructions. Therefore, acquiring and analyzing power consumption data on these devices is less challenging than on a general-purpose computer.

Hoffman et al.~\cite{8} evaluated whether malware can be detected on smartphones by analyzing its power consumption. The proposed approach failed due to the noise introduced by unpredictable user and environment interactions. More recent work by Azmoodeh et al.~\cite{Azmoodeh2017} demonstrated that ransomware can be detected on Android devices by monitoring power consumption alone. 
These works differ from our approach in the way power data was collected. While each used a software that was running on the device being monitored, we use out-of-band hardware monitoring the power profiles on a separate machine. Furthermore, in addition to running non-malicious software and malware separately, we also combine each non-malicious software with malware when collecting the power profiles. 

Power-based integrity assessment for software-defined radio was a 
focus of Gonz\'{a}lez et al.~\cite{gonzalez2009power, gonzalez2011}
who captured the
fine-grained measurements of the processor's power consumption and compared them against signatures from trusted software. This method was adopted by the PFP firm (\url{http://pfpcyber.com/}), which 
characterized the execution of trusted software by extracting its power signatures (patterns that result from the specific sequence of bit transitions during execution) and then using them as a reference to compare time-domain traces in order to determine whether the same code was executing. 
This commercial tool is also applicable to embedded systems~\cite{gonzalez2014detecting, reed2012}. 
It appears that no prior research has tested the efficacy of power profile analysis for malware detection on general-purpose computers. 

Our data analytics method uses multiple anomaly detectors informing a final ensemble detector. 
The use of ensembles of detectors to increase detection accuracy is not uncommon~\cite{bridges2015multi, bridges2016multi, bridges2017setting, christodorescu2007can, harshaw2016graphprints, moore2017modeling}; however, our specific formulation is indeed unique.  
Specific novel contributions include the following:  we introduce a data-driven technique to learn canonical shapes in the power curves (which permits analysis by sequential learning algorithms); 
we test the new Data Smashing~\cite{chattopadhyay2014data} technique for time-series analysis on two separate symbol representations of power data; 
we formulate a z-score, single-feature detector from computing both permutation entropy (see Cao et al.~\cite{cao2004detecting}) and  information variance with application results to power-based malware detection; 
finally, we introduce a simple ``z-score of z-score'' anomaly detection ensemble, and exhibit results showing it outperforms the single-feature detectors lowering false positives but raising recall.

Our detection method uses multiple anomaly detectors informing a final ensemble detector. The strategy is to model heterogeneous characteristics of the power data to increase detection accuracy. This follows previous IDS applications of ensemble anomaly detectors \cite{bridges2015multi, bridges2016multi, christodorescu2007can, harshaw2016graphprints, moore2017modeling}.



\section{Test Bed Design, Rootkits, \& Data Collection}
\label{sec:expDesignDataColl}
Our conjecture is that for a fixed task, malware's actions will produce sufficient deviation in the power profile to admit detection.
To test this, we use three tasks, two tasks that we believe will change in the presence of rootkits\textemdash web browsing with Internet Explorer (IE) and probing the registry\textemdash and a final task, letting the computer sit idle, to test if malware executions are detectable when no other execution commands are given.
IE was chosen because it has been proven that some rootkits affect the performance of browsers~\cite{118}.
The registry was chosen because rootkits often make modifications to it and then use a driver to hide the modified registry keys from observation~\cite{Hoglund08}.
Registry modification is a common element of most Windows programs, so merely watching process behavior for this trait would not give accurate malware detection.
What is uncommon is for programs to then take actions that prevent their registry modifications from being observed by other processes scanning the registry, a concealing characteristic of some malware~\cite{Hoglund08}.


\subsection{Hardware and Software Configuration}
Anticipating that the effects of malware execution on power signals are subtle, significant effort was spent  sensing and recording the power use data.  
Our experiments were run on a Dell OptiPlex 755 machine running 32-bit Windows 7.
To record the data for our experiments, we used a Measurement Computing Data Acquisition\footnote{DAQ model USB-1608G Series with 16 channels used.  \url{www.mccdaq.com}}
(DAQ) system.
Separate channels on the DAQ were connected through a printed circuit board (PCB) sensor to separate voltage sources on the computer's motherboard power connector, and the voltage and current were collected  on each DC power channel.

The vendor-provided DAQ software, TracerDAQ Pro  v. 2.3.1.0,
acquires, analyzes, and displays data from the DAQ's channels~\cite{TracerDAQPro}.
Our experimentation revealed that TracerDAQ Pro admits too little control over sampling rates and experiment length, and that acquired data precision is limited to three decimal places.
Consequently, we developed our own program that directly accesses the DAQ to gather data using custom software that had three distinct advantages: (1) data has 16 bits of precision; 
(2) we were able to better control the sample recording rates and vary the sample timings; (3) we could make real-time calculations that helped assure that data is in the expected time ranges. 

As most malware initiates malicious traffic (e.g., click fraud, beaconing, spamming, etc.), unfiltered Internet  access for the experimental machine is necessary so malware operates with full functionality. For our experiments, an unfiltered, segregated network was created using a cellular data connection to the Internet that passed through another computer, the data repository, that was used to monitor and collect the \begin{wrapfigure}{r}{.28\textwidth}
  \begin{minipage}[b]{0.46\linewidth}
  \centering
    \includegraphics[width=\textwidth,height=1.3in]{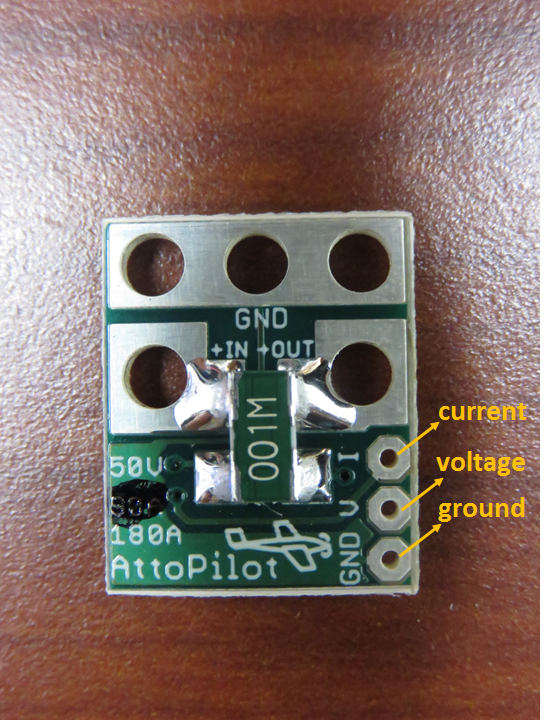}
  \end{minipage}
  ~\hfill~
  \begin{minipage}[b]{0.46\linewidth}
    \centering
      \includegraphics[width=\textwidth,height=1.3in]{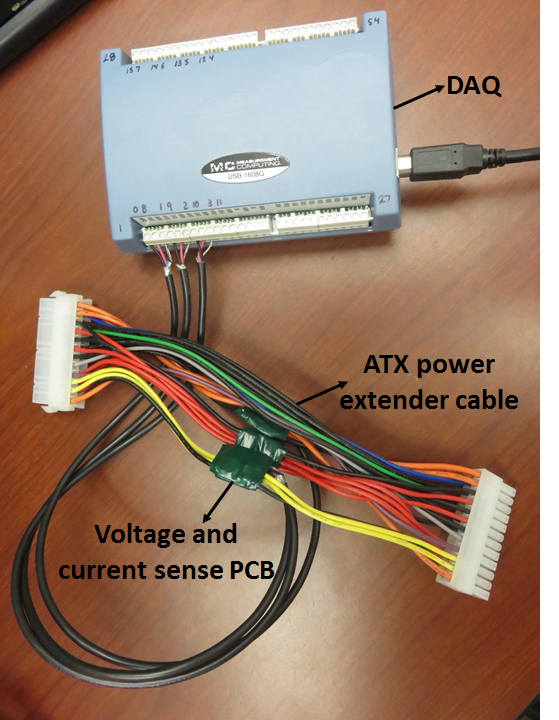}
  \end{minipage}
  \caption{(Left) voltage and current sensor PCB depicted used to obtain the monitor power consumption from the experimental machine.
  (Right) ATX power extender cable attached to the DAQ depicted. The outputs for current, voltage, and ground from the sensor were wired to channels on the DAQ. 
  }
  \label{fig:HW}
\end{wrapfigure}
\noindent power data before connecting to the experimental machine.
Two advantages emerged from this segregated network. (1) This design allowed the rootkits to behave normally, while avoiding any possibility of them infecting other computers on our network. (2) It allowed us to monitor, record, and analyze the experimental machine's network traffic.
For example, using Wireshark, we observed that several rootkits used in our experiments redirected web traffic to advertisement sites and malware as it attempted command and control contact. 
More details about the hardware and software configuration can be found in our technical report~\cite{hernandez2017}.

\subsection{Data Collection \& Rootkits}
\label{sec:data}
\begin{figure*}[!ht]
\includegraphics[width=\textwidth]{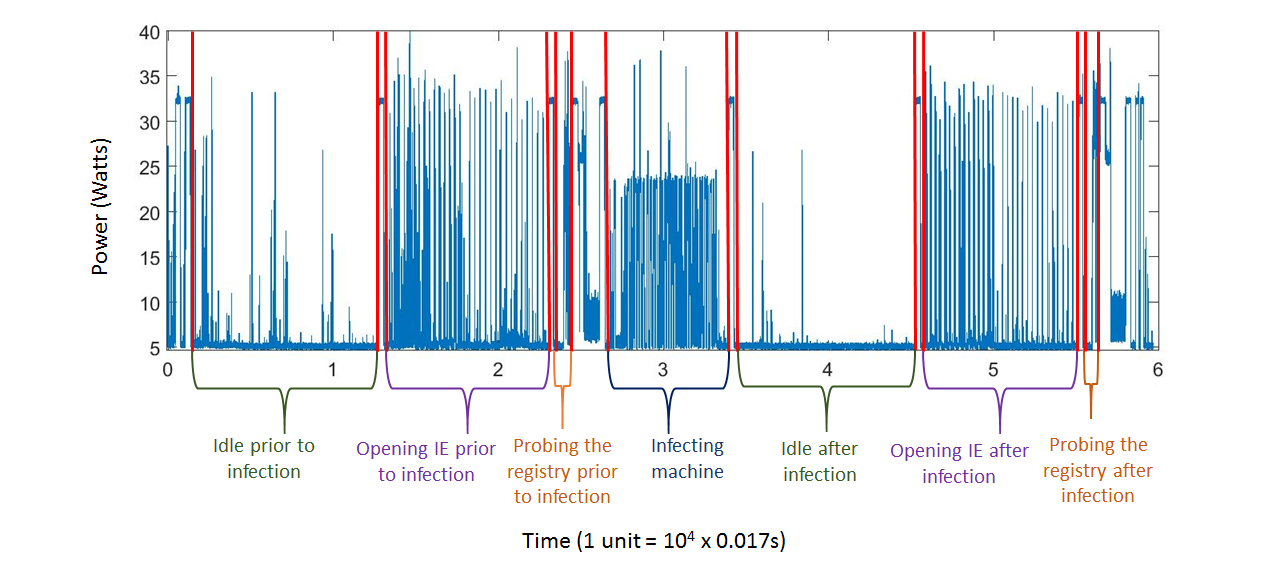}
\centering
\caption{Power profile of +12V CPU depicted for each task before and after infection with Xpaj rootkit. Tasks labeled and demarcated by red vertical bars.}
\label{fig:SeqOfEvents}
\end{figure*}
Although we collected data from eight channels (four voltage channels and four corresponding current channels), the results in this paper only utilize the +12V CPU rail's data.
As part of the data preprocessing, the voltage and current channels were multiplied to obtain a quadruple 
of power consumption for each sample time.
The sampling rate, computed as the median difference in power samples, was 0.0170s, with variance on the order of $10^{-6}$s. 

The power consumption of the general-purpose computer was recorded while completing the idle, IE, and registry tasks, first in a clean state and again after infection. 
No other workload ran on the experimental machine.
To automate the tasks and ensure repeatability between experiments, a Python script orchestrated the following specified sequence of events.
During the data collection process, the experimental machine was idle for a three minute period.
In the case of IE, fifteen windows were open with a five second delay between each, and then each of these windows was closed before the script continued with the next task, probing the registry.
In case of probing the registry, we used regedit to copy all the information from the Windows registry to a .reg file.
By scripting workflow tasks, we ensure a repeatable sequence of commands.
To delimit and identify tasks in the power profile, a  C++ program was written and called between each task in order to insert ``markers'' that could be recorded by the DAQ.
This program performed calculations stressing the CPU of the experimental machine to 100\%, causing a noticeable plateau in the CPU power profile.
After data collection, a MATLAB script processed the power profile data, detected these embedded markers as start and end points to distinguish between different tasks of the experiment.
This is the source of the data segments used in the malware detection algorithms presented in this paper.

To ensure the experimental machine was returned to the same initial state between experiments, we made a clean installation of Windows using Clonezilla (\url{www.clonezilla.org/}). 
With this preparation, the Python script was started on the experimental machine, while power consumption data was recorded on the data repository.
After completing the sequence of tasks in this uninfected state, the experimental machine was infected with a particular rootkit.
Then the sequence of tasks was repeated by the Python script and, after collection, the power consumption data was labeled as infected.
Fig. \ref{fig:SeqOfEvents} shows  power data for the +12V CPU rail from a single run of the Python script.

As seen in Fig. \ref{fig:SeqOfEvents}, even within the sections where the machine is idle, there are still periods of elevated power use.
We investigated these events by correlating elevated power use events to Windows processes starting and ending.
We were able to attribute these power use events to Microsoft-provided background services which are a normal part of the Windows environment.
This illustrates the main challenge of profiling power consumption of fixed tasks 
on a general-purpose computer, background processes contribute noise to the power profiles.
The hypothesis under investigation is that the executions of malware during various tasks and idle periods produce a significant-enough change in the power usage to admit accurate detection in the presence of the ambient but somewhat unpredictable normal executions.


Next we discuss the rootkit samples used.
A rootkit is a ``kit'' consisting of small programs that allow a permanent or consistent, undetectable presence on a computer~\cite{Hoglund08}.
Basically, a rootkit locates and modifies the software on the computer system
with the purpose of masking its behavior; for example, patching, a technique that modifies the data bytes encoded in an executable code, is an example of a type of modification that can be made~\cite{Hoglund08}.

For our experiments, five different malware programs were used, and
each exhibits traits of a rootkit and/or actions for camouflaging itself.
Alureon, also known as TDL4 or TDSS, is a Trojan that attempts to steal personal data.
It encrypts itself in the master boot record, which makes it exceptionally hard to detect~\cite{ref:alureon1}, can remove competitor's malware, and hides communication via encryption. 
Pihar, an Alueron variant also known as Purple Haze, is capable of changing system settings without any manual direction, reconfiguring the Windows registry editor or keys, and disabling the AV software~\cite{ref:Pihar1}.
Like Alureon and Pihar, Xpaj is another rootkit that hides in the master boot record~\cite{ref:Xpaj1}. 
Xpaj uses several methods of code obfuscation and encryption to hide its presence in the infected file.
On the other hand, Sirefef, also known as Zero Access, is a trojan that is used to download other malware on the infected machine from a botnet that is mostly involved in bitcoin mining and click fraud.
Sirefef also overrides the registry and browser settings to hide itself~\cite{ref:ZeroAccess1}.
Like Alureon, it will spoof online search queries and the user will be redirected to malicious websites.
Finally, Max rootkit is a Trojan that sometimes destroys information and files on its host computer, and changes Windows Explorer settings to download other malicious files~\cite{ref:MaxRootkit1}.

Upon installation, to verify that the experimental machine was successfully infected with each one of these rootkit examples, two malware removal tools (TDSSkiller~\cite{ref:TDSSkiller} and Microsoft Windows Malicious Software Removal Tool~\cite{ref:WindowsRootkitRemovalTool}) were used.  
 Note that these tools were not used to clean the experimental machine.
We analyzed each rootkit using the web-based application called VxStream  Sandbox~\cite{ref:VxStreamSandbox}
 to see which Windows API functions were being called.
In total, for each of the three tasks (Idle, IE, and Registry) we collected power profile data for 15 runs in an uninfected state and 15 in an infected state (three per rootkit).



\section{Data Analysis Approach \& Results}
\label{sec:analysis}
Exploring the hypothesis that malware execution necessarily changes the power profile, we formulate and test two detection approaches.
The first approach is unsupervised, viewing the detection task as a one-class anomaly detection problem.
This gives us slightly stronger initial hypotheses than other anomaly detection applications, which see all historical data out of which only a small portion are positive (malicious/noteworthy) and the majority negative (normal).
For example, see Harshaw et al.~\cite{harshaw2016graphprints} where robust statistic methods were used to accommodate possible attacks in historical network data by preventing outlier observations (possibly unknown attack) from affecting their models.
Our unsupervised detection method proceeds by creating a baseline from ten uninfected power profiles and detecting sufficient deviations in test profiles as infected.
This has the obvious advantage that no prior observations of data collected under the presence of malware is necessary.
The second, supervised approach assumes knowledge of previously observed uninfected and infected data.
More specifically, we train and test on an equal number of uninfected and infected profiles, but the positively-labeled testing profiles are those  rootkit-infected profiles never seen in training.
Hence, both scenarios are designed to test detection capabilities against never-seen-before malware.
Both use the same set of features.

\subsection{Features}
\label{sec:features}
Modeling diverse qualities of the data gives more avenues of detection; hence, we craft a variety of features modeling statistical properties of the power data sample as well as many time-varying characteristics.
For each task (IE, Registry, Idle), the power profile of a given run is transformed into a feature vector.
The first four features are statistical moments, which encode characteristics of the power data regarded as a set of independent and identically distributed (IID) samples of a distribution.
The remaining features ($L^2$ distance, permutation entropy, and Data Smashing distances) seek to describe time-dependent properties of the power profile.
For these features it is convenient to have uniformly sampled power data.
To obtain uniform times, all observations in a run are shifted by their initial time, so time uniformly begins at $t=0$.
Next, the time-stamped data points are linearly connected to define the power profile for all time values, and then sampled for each multiple of $.01$ seconds.
Note power samples occur with a median $.0170$ seconds apart and variance on the order of $10^{-6}$ seconds, so by using the finer-grained samples, we obtain a close approximation of the observed profile that is also uniformly spaced.

\subsubsection{Statistical Moments}
\label{sec:moments}
In order to model the distribution of the power data, we use (empirical estimates of) the mean (first moment), variance, skewness, and kurtosis (second-fourth normalized central moments).\footnote{Kurtosis $:=  \sum_{i = 1}^N (x_i - \mu)^4/(N\sigma^4) - 3.$ This follows Fisher's definition, with 3 subtracted so normally distributed data yield 0.}

\begin{figure*}
    \includegraphics[scale=0.5]{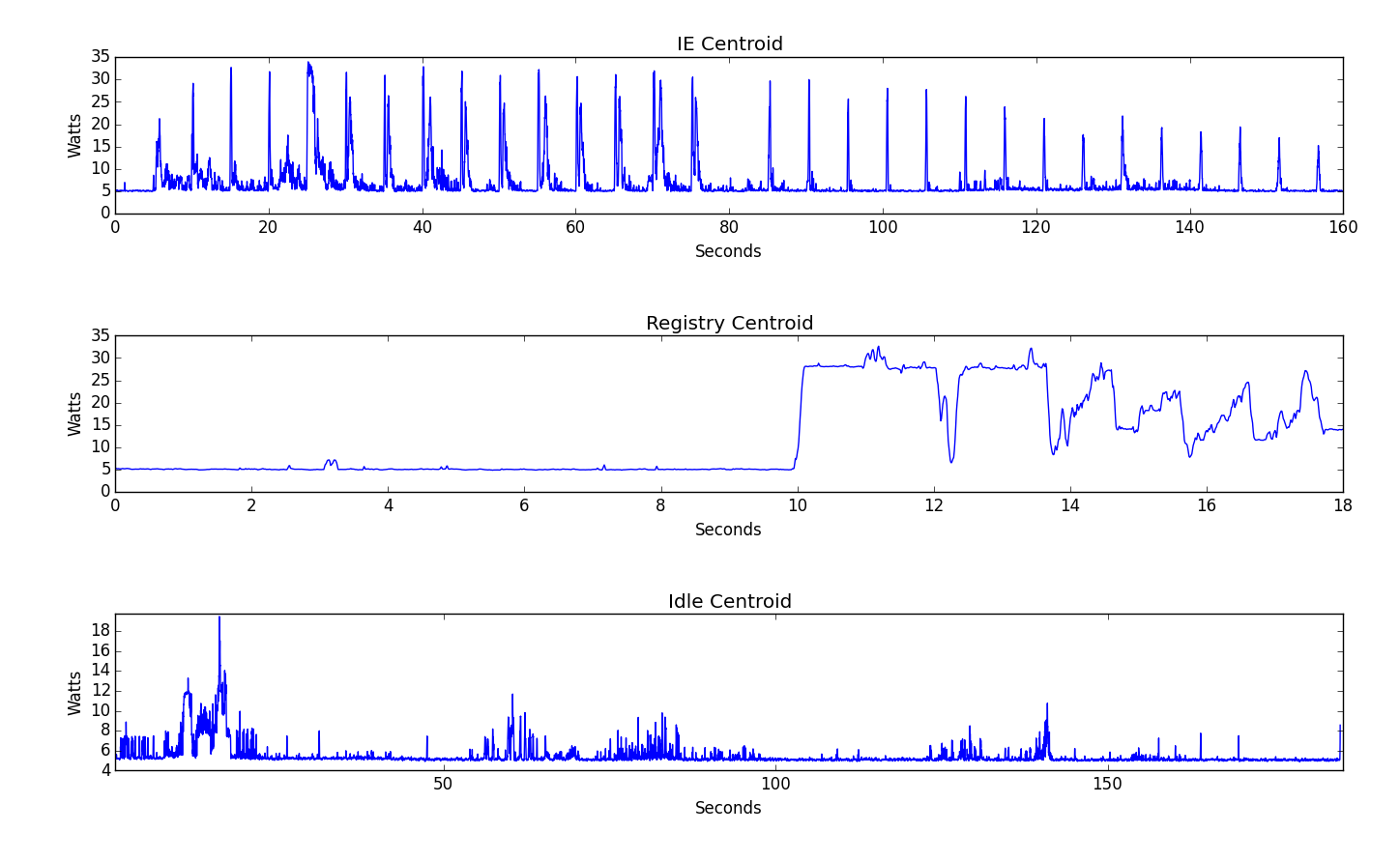}
    \centering
    \caption{(Watts vs. Seconds) Canonical profiles (baselines for $L^2$ error) per task found by clustering unsupervised training data (ten uninfected runs) with $k$-means clustering with $L^2$ distance. Gap statistic used to find the no. of clusters, $k$, and $k = 1$ (as expected) is found in each case. Centriods (means of each cluster) depicted.
    }
    \label{fig:centroids}
\end{figure*}
\subsubsection{$L^2$-Norm Error}
\label{sec:l2}
Regarding each power profile as a function $p(t)$, the $L^2$ norm (defined as $[ \int |p|^2 dt]^{1/2}$)  furnishes a natural (Euclidean) distance for pairwise comparison.
As we have constructed uniformly sampled approximations of the profiles, multiplying the usual ($l^2$ Euclidean) vector distance to the tuple of power observations by the $\Delta t = .01$ gives a close approximation to the integral.
To construct baseline profile, we cluster the uninfected training data using $k-$means clustering, with $k$ determined by the gap statistic method~\cite{tibshirani2001estimating}.
This provides a sanity check\textemdash we hypothesize that our uninfected training runs should be similar; hence, we expect a single cluster with the centroid a canonical ``baseline'' function.
Indeed, for all tasks a single canonical baseline was found by the gap statistic method.
Fig.~\ref{fig:centroids} depicts the baselines for each task, as found on the ten uninfected training runs used for unsupervised training.
For the supervised approach, the baseline is created from the 12 clean training profiles.
Finally, the feature for a given profile is the $L^2$ distance to the baseline.

\subsubsection{Permutation Entropy}
\label{sec:pe}
Permutation Entropy, is a method of time-series analysis that detects deviations in the short-term wiggliness (roughly speaking) and has been used in various time-series detection applications including  epilepsy~\cite{cao2004detecting} and motor bearing fault detection~\cite{zhang2015novel}.
Our formulation follows Cao et al.~\cite{cao2004detecting}.
From a time-series data vector, $(x_1, \dots, x_N)$ (a power profile in our case) fix an $m$-length contiguous sub-vector $(x_i, x_{i+1}, ..., x_{i+m-1})$ and denote this vector as $(y_0, ..., y_{m-1})$.
Next sort the elements into increasing order, $(y_{j_0}, ..., y_{j_{m-1}})$, so either $y_{j_k} < y_{j_{k+1}}$  or $y_{j_k} = y_{j_{k+1}}$ and $j_k + 1$ = $j_{k+1}$.
Now, $(j_0, ..., j_{m-1})$ is the permutation of the numbers $(0, 1,..., m-1)$ indicating the rearrangement necessary to sort the $m$-length vector.
Given a time-series data vector $(x_1, \dots, x_n)$ (a power profile), we extract contiguous sub-vectors of length $m$ with overlap $m/2$, i.e., $(x_1, ..., x_m)$, $(x_{\frac{m}{2}+1}, ... , x_{m + \frac{m}{2}}), $ etc.
Each $m$-length vector is converted to one of the $m!$ permutations by the process above, and we represent the time-series as a bag of permutations.
Our goal is to learn how rare/likely the profile is by understanding the likelihood of its permutations.
To do this, we require a probability distribution over the sample space of $m!$ permutations.

For a fixed task, we estimate the needed probability distribution from the training runs on that task.
Let $\mathcal{O}$ be the observations (multiset) of permutations across all training runs, and for each permutation $\gamma = (j_0, ..., j_{m-1})$,
let $\#( \gamma , \mathcal{O} )$ denote the number of times $\gamma$ is observed in $\mathcal{O}$.
Then we use the maximum a posteriori (MAP) estimation with uniform prior to estimate $\gamma$'s probability,
$P( \gamma ) :=  (\#( \gamma, \mathcal{O} ) +1)/(|\mathcal{O}|+m!).$
This allows permutations never observed in training to have positive likelihood, albeit very small.
It is clear from the MAP estimate that $m$ must be chosen carefully, as for moderate values of $m$, $m!$ will easily dominate $|\mathcal{O}|$, washing out the contribution of our observations.
We chose $m = 6$, which gives $m! = 720$, while $|\mathcal{O}|$ = 61,680 for IE, 6,240 for Registry, and 53,980 for Idle tasks.
Finally, for a given run we will compute the entropy of the observed permutations as IID samples from the distribution above.
After converting an observed power profile to a bag of permutations, $(\gamma_i: i = i, ..., n )$, we compute the permutation entropy, that is, the expected information of the permutation,
$\hat{H} = (1/n) \sum_i -\log(P(\gamma_i))P( \gamma_i ).$

\subsubsection{Data Smashing Distances}
\label{sec:dsd}
Data Smashing distance (DSD) is a new algorithm pioneered by Chattopadhyay and  Lipson~\cite{ chattopadhyay2014data} 
for quantifying the distance between two time-varying sequences.
DSD assumes the non-restrictive hypothesis that each sequence is a sample from a probabilistic finite state automata (PFSA) and approximates the distance between the underlying PFSAs.\footnote{A PFSA is (roughly speaking) a generative model for sampling a sequence of symbols and is composed of a set of states, a stochastic state transition matrix, and rules giving fixed symbol emissions on the transitions. PFSA are similar to, but conceptually simpler than, hidden markov models (HMMs). HMMs emit a sequence of symbols in a two-step process. The new state is sampled according to the transition probabilities. Then a symbol is sampled from the emission distribution for the new state. Rather than a probabilistic symbol emission depending on the state, PFSA emit a fixed symbol determined by the transition.}
Roughly speaking, PFSA are applicable models to any sequence of symbols where the probability of a current symbol depends only on the previous symbols.
The DSD algorithm relies on mathematical developments involving PFSA by Wen and Ray~\cite{wen2012vector} who discover a groups structure and metric on the set of PFSA. 
DSD gives a novel, computationally efficient algorithm for approximating the distance between two PFSA from sufficiently long symbol strings sampled from each.
This operationalizes the developments of Wen and Ray, as one need not know nor try to infer the PFSA from the symbol observations.

In order to apply DSD, users must transform their time-varying data into a sequence of symbols that is sufficiently long.
The expected string length required by the DSD algorithm is strictly increasing in the number of symbols (factorial in this variable) and the product of the symbol probabilities.\footnote{For details see the Proposition 13 in the supplementary text of
~\cite{chattopadhyay2014data}.}
Although the desired PFSA distance is symmetric ($d(s1, s2) = d(s2, s1)$), DSD uses a probabilistic algorithm to approximate the distance, and will produce similar but, in general, not identical values as the order is switched.
For this reason, given two strings $s_1$ and $s_2$, we compute $d(s1, s2)$ and $d(s2, s1)$ and store the average of the two as the distance.
If either distance is not returned, it indicates our strings are not long enough, and this feature is not used.

We employ DSD on two symbol representations of the power data, separately.
The first transforms power samples into two symbols, ``high'' and ``low'', and is a stereotypical application of Data Smashing.
This is referred to hereafter as ``DSD Data Distance.''
The threshold for the binary representation is chosen for each pair of profiles, so the number of power readings over and under the threshold are equal.
This maximizes the product of the probabilities of each symbol, which helps ensure the symbols are sufficiently long.
Note that the baseline observed profiles are fixed, and to compute the DSD Data Distance between the baseline and a newly observed given test profile, the threshold for encoding both profiles as a binary high/low sequence is dependent on both profiles.
Hence, the binary sequence for the baseline profile will be slightly different in each comparison, but no prior knowledge of any test data is necessary.
\begin{wrapfigure}{r}{.50\linewidth}
    \includegraphics[scale=0.23]{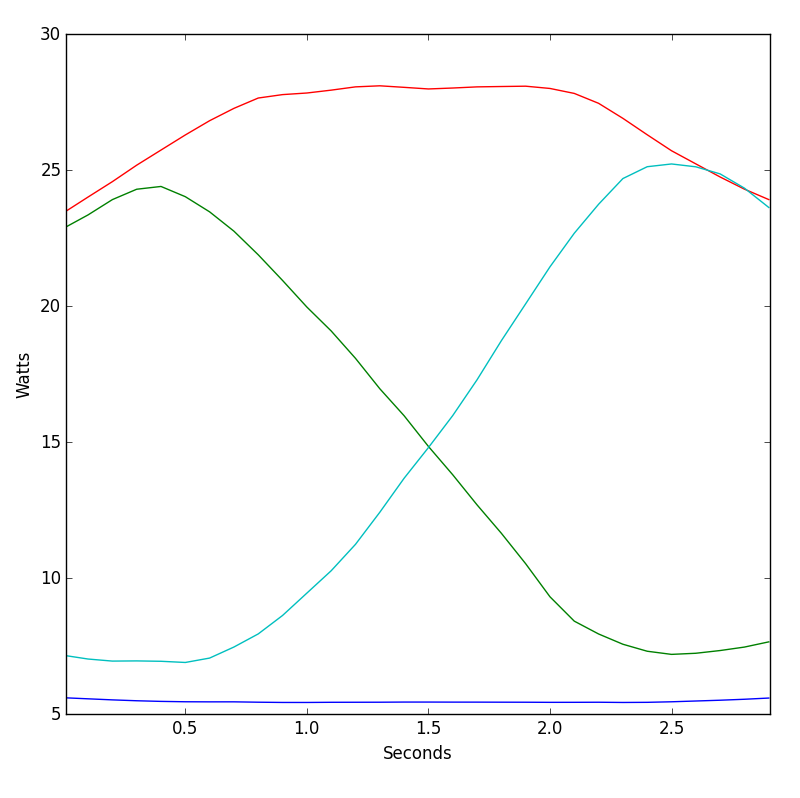}
    \centering
    \caption{(Watts vs. Seconds) Canonical shapes found by clustering three-second intervals of training data. Note that the four shapes characterize {\color{red}high power}, {\color{blue} low power}, {\color{darkolivegreen}  falling power}, and {\color{teal} rising power}. }
    \label{fig:shapes}
\end{wrapfigure}

The second application seeks to model sequences of shapes in the power curves and is referred to as ``DSD Shape Distance.''
To produce the shapes, the training profiles are cut into three-second intervals with 1.5-second overlap.
Then the bag of all three second snippets are clustered using $k$-means and $L^2$ distance as in Sec.~\ref{sec:l2}.
We manually investigated the centroids as $k$ varied, choosing $k=4$ as is maximal for which the centroids are not very similar curves.
See Fig.~\ref{fig:shapes} depicting the four canonical shapes learned from the data.
Finally, given a power profile, we cut it into a sequence of three-second intervals, overlapping by 1.5 seconds, and assign to each interval
the closest canonical shape.
This transforms a power profile into a sequence of four ``shape'' symbols.
Our conjecture is that profiles that exhibit a clear pattern when uninfected (e.g., the IE baseline, top Fig.~\ref{fig:centroids}), will give a regular sequence of shapes that may be disrupted by malware execution.

In both transformations to a sequence of symbols, every sequence is concatenated with itself 100 times to ensure the strings are long enough to perform the algorithm.
This can be conceptualized as a simulation of running the same task repeatedly 100 times.
DSD Shape distance is not applicable to the the Idle task runs because the strings are not long enough.
This is likely caused by the fact that the flat, low power shape dominates the probabilities, making the product of the four symbols probabilities very small.
To obtain both the DSD Data and DSD Shape feature value for a given a power profile, we compute the distance between its symbol sequence and that of all ten training runs concatenated together.

\subsection{Unsupervised Approach: Anomaly Detection Ensemble}
\label{sec:unsupervised}
Treating the malware detection task as a one-class problem, we construct a baseline for each feature from ten uninfected 
 \newgeometry{left=0.75 in,right=0.75 in,bottom=0.75 in,top=0.75 in} 
\begin{landscape}
\centering
\begin{threeparttable}
   \centering 
    \caption{Unsupervised Test Data Z-Scores \& Detection Results\tnote{1}}
    {\scriptsize
    \begin{tabular}{clccccccccccccccccccccccc}
    \toprule
           &  & \multicolumn{5}{c}{\textcolor[rgb]{ .122,  .286,  .49}{Uninfected}} & \multicolumn{3}{c}{\textcolor[rgb]{ .502,  0,  0}{Alureon}} & \multicolumn{3}{c}{\textcolor[rgb]{ .502,  0,  0}{Max Rootkit}} & \multicolumn{3}{c}{\textcolor[rgb]{ .502,  0,  0}{Pihar}} & \multicolumn{3}{c}{\textcolor[rgb]{ .502,  0,  0}{Sirefef}} & \multicolumn{3}{c}{\textcolor[rgb]{ .502,  0,  0}{Xpaj}} & \multicolumn{1}{l}{Threshold} & \multicolumn{1}{l}{TPR} & \multicolumn{1}{l}{FDR} \\
    \cmidrule(lr){1-2}
    \cmidrule(lr){3-7}
    \cmidrule(lr){8-10}
    \cmidrule(lr){11-13}
    \cmidrule(lr){14-16}
    \cmidrule(lr){17-19}
    \cmidrule(lr){20-22} 
    \cmidrule(lr){23-25}
    
          & \textbf{Feature/Run}   & 1     & 2     & 3     & 4     & 5     & 1     & 2     & 3     & 1     & 2     & 3     & 1     & 2     & 3     & 1     & 2     & 3     & 1     & 2     & 3    \\
    \parbox[c]{4mm}{\multirow{3}{*}{\rotatebox[origin=c]{90}{\textbf{IE Task}\phantom{dgldgl}}}} & Mean  & 0.53  & \textbf{1.13} & 0.59  & 0.95  & 0.96  & \textbf{4.40} & \textbf{5.26} & \textbf{2.52} & \textbf{3.33} & \textbf{3.29} & \textbf{3.68} & \textbf{2.10} & \textbf{2.30} & \textbf{2.06} & \textbf{1.38} & \textbf{3.59} & \textbf{3.43} & \textbf{2.06} & \textbf{2.19} & \textbf{2.68} &       & 1.00  & 0.06 \\
          & Variance & 0.55  & 0.40  & \textbf{1.09} & 0.46  & 0.46  & \textbf{4.56} & \textbf{5.52} & \textbf{1.83} & \textbf{2.88} & \textbf{2.59} & \textbf{2.48} & \textbf{2.38} & \textbf{2.20} & \textbf{1.56} & \textbf{1.56} & \textbf{3.22} & \textbf{2.39} & \textbf{1.89} & \textbf{2.00} & \textbf{1.80} &       & 1.00  & 0.06 \\
          & Skewness & 0.55  & 0.40  & \textbf{1.09} & 0.46  & 0.46  & \textbf{4.56} & \textbf{5.52} & \textbf{1.83} & \textbf{2.88} & \textbf{2.59} & \textbf{2.48} & \textbf{2.38} & \textbf{2.20} & \textbf{1.56} & \textbf{1.56} & \textbf{3.22} & \textbf{2.39} & \textbf{1.89} & \textbf{2.00} & \textbf{1.80} &       & 1.00  & 0.06 \\
          & Kurtosis & 0.55  & 0.40  & \textbf{1.09} & 0.46  & 0.46  & \textbf{4.56}  & \textbf{5.52} & \textbf{1.83} & \textbf{2.88} & \textbf{2.59} & \textbf{2.48} & \textbf{2.38} & \textbf{2.20} & \textbf{1.56} & \textbf{1.56} & \textbf{3.22} & \textbf{2.39} & \textbf{1.89} & \textbf{2.00} & \textbf{1.80} &       & 1.00  & 0.06 \\
          & $L^2$ Error & 0.88  & 0.72  & 0.61  & 0.02  & 0.31  & \textbf{2.55} & \textbf{2.35} & \textbf{2.63} & \textbf{1.34} & \textbf{2.07} & \textbf{1.42} & \textbf{2.66} & \textbf{1.62} & \textbf{1.96} & \textbf{2.04} & \textbf{2.04} & \textbf{1.09} & \textbf{1.29} & \textbf{1.09} & \textbf{1.68} &       & 1.00  & 0.00 \\
          & Perm. Entropy\tnote{4} & 0.02  & 0.04  & 0.01  & 0.06  & 0.06  & 0.05  & 0.01  & 0.07  & 0.01  & 0.05  & 0.09  & 0.08  & 0.08  & 0.03  & 0.03  & 0.00  & 0.04  & 0.02  & 0.02  & 0.07  &       & 0.00  & 0.00 \\
          & DSD (Data)\tnote{5}  & \textbf{1.32} & \textbf{2.94} & 0.38  & \textbf{2.46} & \textbf{1.94} & \textbf{3.91} & \textbf{3.66} & \textbf{3.79} & \textbf{3.37} & \textbf{4.35} & \textbf{5.59} & 0.93  & \textbf{1.45} & \textbf{3.13} & 0.85  & \textbf{3.99} & \textbf{5.61} & \textbf{1.72} & \textbf{2.27} & \textbf{4.20} &       & 0.87  & 0.24 \\
          & DSD (Shape)\tnote{5} & \textbf{1.52} & \textbf{1.25} & 0.55  & 0.23  & \textbf{2.08} & \textbf{1.00} & \textbf{7.61} & 0.44  & \textbf{2.60} & \textbf{1.32} & 0.52  & 0.59  & 0.66  & \textbf{1.55} & \textbf{1.05} & 0.78  & 0.72  & \textbf{2.03} & \textbf{2.21} & 0.56  &       & 0.53  & 0.27 \\ 
          & IE Votes\tnote{3} & 2     & 3     & 3     & 1     & 2     & {\textbf{7}} & {\textbf{7}} & {\textbf{6}} & {\textbf{7}} & {\textbf{7}} & {\textbf{6}} & {\textbf{5}} & {\textbf{6}} & {\textbf{7}} & {\textbf{6}} & {\textbf{6}} & {\textbf{6}} & {\textbf{7}} & {\textbf{7}} & {\textbf{6}} & 4.7   & 1.00  & 0.00 \\
    \cmidrule(lr){1-2}
    \cmidrule(lr){3-7}
    \cmidrule(lr){8-10}
    \cmidrule(lr){11-13}
    \cmidrule(lr){14-16}
    \cmidrule(lr){17-19}
    \cmidrule(lr){20-22} 
    \cmidrule(lr){23-25}
 
  \parbox[c]{4mm}{\multirow{3}{*}{\rotatebox[origin=c]{90}{\textbf{Registry Task \phantom{dgl}} }}}  & Mean  & 0.94  & \textbf{1.11} & \textbf{1.22} & \textbf{1.52} & 0.19  & \textbf{1.28} & \textbf{1.89} & \textbf{1.27} & \textbf{2.19} & \textbf{1.96} & \textbf{1.99} & \textbf{1.08} & \textbf{2.22} & \textbf{1.32} & \textbf{1.21} & \textbf{1.84} & \textbf{2.00} & \textbf{1.88} & \textbf{2.07} & \textbf{2.06} &       & 1.00  & 0.17 \\
          & Variance & \textbf{1.01} & \textbf{1.05} & \textbf{1.24} & \textbf{1.48} & 0.00  & \textbf{1.09} & \textbf{1.95} & \textbf{1.06} & \textbf{1.96} & \textbf{2.04} & \textbf{2.10} & \textbf{1.05} & \textbf{2.03} & \textbf{1.27} & \textbf{1.41} & \textbf{2.23} & \textbf{2.13} & \textbf{1.81} & \textbf{2.41} & \textbf{2.04} &       & 1.00  & 0.21 \\
          & Skewness & \textbf{1.01} & \textbf{1.05} & \textbf{1.24} & \textbf{1.48} & 0.00  & \textbf{1.09} & \textbf{1.95} & \textbf{1.06} & \textbf{1.96} & \textbf{2.04} & \textbf{2.10} & \textbf{1.05} & \textbf{2.03} & \textbf{1.27} & \textbf{1.41} & \textbf{2.23} & \textbf{2.13} & \textbf{1.81} & \textbf{2.41} & \textbf{2.04} &       & 1.00  & 0.21 \\
          & Kurtosis & \textbf{1.01} & \textbf{1.05} & \textbf{1.24} & \textbf{1.48} & 0.00  & \textbf{1.09} & \textbf{1.95} & \textbf{1.06} & \textbf{1.96} & \textbf{2.04} & \textbf{2.10} & \textbf{1.05} & \textbf{2.03} & \textbf{1.27} & \textbf{1.41} & \textbf{2.23} & \textbf{2.13} & \textbf{1.81} & \textbf{2.41} & \textbf{2.04} &       & 1.00  & 0.21 \\
          & $L^2$ Error & 0.93  & \textbf{1.20} & \textbf{1.00}  & 0.89  & \textbf{1.55 } & \textbf{2.02} & \textbf{2.39} & \textbf{1.48} & \textbf{2.22} & \textbf{2.52} & \textbf{2.27} & \textbf{1.71} & \textbf{2.72} & \textbf{2.05} & \textbf{1.69} & \textbf{2.50} & \textbf{2.32} & \textbf{2.42} & \textbf{2.63} & \textbf{2.03} &       & 1.00  & 0.12 \\
          & Perm. Entropy\tnote{4} & 0.02  & 0.17  & 0.15  & 0.04  & 0.16  & 0.11  & 0.06  & 0.09  & 0.15  & 0.17  & 0.17  & 0.21  & 0.19  & 0.13  & 0.10  & 0.14  & 0.11  & 0.17  & 0.09  & 0.11  &       & 0.00  & 0.00 \\
          & DSD (Data)\tnote{5}  & 0.69  & 0.68  & 1.43  & 0.86  & 0.05  & \textbf{1.57} & \textbf{1.20} & \textbf{2.73} & 0.78  & 0.90  & \textbf{1.08} & \textbf{1.18} & 0.76  & \textbf{1.47} & \textbf{1.00} & \textbf{1.04} & 0.78  & \textbf{1.58} & \textbf{1.12} & \textbf{1.13} &       & 0.67  & 0.09 \\
          & DSD (Shape)\tnote{5} & 0.42  & 0.39  & 0.06  & \textbf{1.73} & 0.18  & 0.07  & \textbf{6.21} & 0.78  & \textbf{5.68} & \textbf{6.55} & \textbf{5.62} & 0.04  & \textbf{4.23} & 0.05  & 0.28  & \textbf{5.63} & \textbf{6.40} & \textbf{4.64} & \textbf{5.29} & \textbf{5.91} &       & 0.67  & 0.09 \\
          & Registry  Votes\tnote{3} & 3     & 5     & 5     & 5     & 1     & {\textbf{6}} & {\textbf{7}} & {\textbf{6}} & {\textbf{6}} & {\textbf{6}} &{\textbf{7}} & {\textbf{6}} & {\textbf{6}} & {\textbf{6}} & \textbf{6}     & {\textbf{7}} & {\textbf{6}} & {\textbf{7}} & {\textbf{7}} & {\textbf{7}} & 5.2   & 1.00  & 0.00 \\
       \cmidrule(lr){1-2}
    \cmidrule(lr){3-7}
    \cmidrule(lr){8-10}
    \cmidrule(lr){11-13}
    \cmidrule(lr){14-16}
    \cmidrule(lr){17-19}
    \cmidrule(lr){20-22} 
    \cmidrule(lr){23-25}
 
         & Mean  & \textbf{1.63} & \textbf{1.43} & 0.85  & \textbf{2.26} & \textbf{2.00} & \textbf{6.05} & \textbf{5.58} & \textbf{5.55} & \textbf{4.61} & \textbf{6.69} & \textbf{7.83} & \textbf{4.08} & \textbf{4.59} & \textbf{4.76} & \textbf{3.85} & \textbf{6.30} & \textbf{7.16} & \textbf{5.33} & \textbf{5.55} & \textbf{5.67} &       & 1.00  & 0.21 \\
    \parbox[c]{4mm}{\multirow{3}{*}{\rotatebox[origin=c]{90}{\textbf{Idle Task \phantom{l}} }}}    & Variance & 0.55  & 0.46  & 0.04  & \textbf{1.94} & 0.48  & \textbf{7.83} & \textbf{7.51} & \textbf{4.20} & \textbf{5.29} & \textbf{7.98} & \textbf{8.04} & \textbf{6.99} & \textbf{7.04} & \textbf{5.10} & \textbf{4.70} & \textbf{8.08} & \textbf{7.50} & \textbf{7.98} & \textbf{8.05} & \textbf{5.37} &       & 1.00  & 0.06 \\
          & Skewness & 0.55  & 0.46  & 0.04  & \textbf{1.94} & 0.48  & \textbf{7.83} & \textbf{7.51} & \textbf{4.20} & \textbf{5.29} & \textbf{7.98} & \textbf{8.04} & \textbf{6.99} & \textbf{7.04} & \textbf{5.10} & \textbf{4.70} & \textbf{8.08} & \textbf{7.50} & \textbf{7.98} & \textbf{8.05} & \textbf{5.37} &       & 1.00  & 0.06 \\
          & Kurtosis & 0.55  & 0.46  & 0.04  & \textbf{1.94} & 0.48  & \textbf{7.83} & \textbf{7.51} & \textbf{4.20} & \textbf{5.29} & \textbf{7.98} & \textbf{8.04} & \textbf{6.99} & \textbf{7.04} & \textbf{5.10} & \textbf{4.70} & \textbf{8.08} & \textbf{7.50} & \textbf{7.98} & \textbf{8.05} & \textbf{5.37} &       & 1.00  & 0.06 \\
          & $L^2$ Error & \textbf{2.02} & \textbf{2.67} & \textbf{2.11} & 0.51  & 0.96  & \textbf{6.14} & \textbf{5.62} & 0.02  & \textbf{1.46} & \textbf{6.20} & \textbf{6.10} & \textbf{4.76} & \textbf{4.80} & \textbf{1.66} & 0.74  & \textbf{6.32} & \textbf{5.44} & \textbf{6.37} & \textbf{6.37} & \textbf{1.78} &       & 0.87  & 0.19 \\
          & Perm. Entropy\tnote{4} & 0.03  & 0.05  & 0.03  & 0.03  & 0.07  & 0.02  & 0.00  & 0.05  & 0.02  & 0.03  & 0.09  & 0.01  & 0.08  & 0.02  & 0.01  & 0.02  & 0.06  & 0.03  & 0.04  & 0.09  &       & 0.00  & 0.00 \\
          & DSD (Data)\tnote{5}  & \textbf{1.88} & \textbf{2.31} & \textbf{1.00} & \textbf{1.78} & \textbf{1.94} & \textbf{2.13} & \textbf{1.78} & \textbf{3.46} & \textbf{1.77} & \textbf{2.75} & \textbf{3.71} & \textbf{0.13} & \textbf{0.89} & \textbf{2.01} & \textbf{1.55} & \textbf{2.25} & \textbf{3.37} & \textbf{1.12} & \textbf{1.47} & \textbf{2.52} &       & 0.87  & 0.24 \\
          & Idle Votes\tnote{3} & 3     & 3     & 2     & {\textbf{5}} & 2     & {\textbf{6}} &  {\textbf{6}} &  {\textbf{5}} &  {\textbf{6}} &  {\textbf{6}} &  {\textbf{6}} &  {\textbf{5}} &  {\textbf{5}} &  {\textbf{6}} &  {\textbf{5}} & {\textbf{6}} & {\textbf{6}} & {\textbf{6}} & {\textbf{6}} & {\textbf{6}} & 3.9   & 1.00  & 0.06 \\
      \cmidrule(lr){1-2}
    \cmidrule(lr){3-7}
    \cmidrule(lr){8-10}
    \cmidrule(lr){11-13}
    \cmidrule(lr){14-16}
    \cmidrule(lr){17-19}
    \cmidrule(lr){20-22} 
    \cmidrule(lr){23-25}
 
    \multicolumn{2}{l}{\textbf{Total Votes\tnote{2}}}  
    & 8  & 11 & 10 & 11 & 5 
    & \cellcolor{lightred}{\textbf{19}} & \cellcolor{lightred}{\textbf{20}} & \cellcolor{lightred}{\textbf{17}} & \cellcolor{lightred}{\textbf{19}} & \cellcolor{lightred}{\textbf{19}} & \cellcolor{lightred}{\textbf{19}} & \cellcolor{lightred}{\textbf{16}} & \cellcolor{lightred}{\textbf{17}} & \cellcolor{lightred}{\textbf{19}} & \cellcolor{lightred}{\textbf{16}} & \cellcolor{lightred}{\textbf{19}} & \cellcolor{lightred}{\textbf{18}} & \cellcolor{lightred}{\textbf{20}} & \cellcolor{lightred}{\textbf{20}} & \cellcolor{lightred}{\textbf{19}}  & 11.2  & 1.0 & 0.0 \\
    \bottomrule 
    \end{tabular}%
    } 
    \label{tab:unsupervised-testing}%
\begin{tablenotes}[para]
\small
\item[1] Table depicts z-scores (computed using training features' means and variances) of test set\textemdash five clean runs and three runs of each rootkit for all tasks. 
A z-score at least one constitutes and anomalous vote, and these values are depicted in bold font. 
Threshold column gives the minimum number of votes needed for positive classification, computed as the mean plus one standard deviation of the training vote totals. 
Sums of votes for each task reported, with task totals over threshold also in bold. 
Anomalous classification as informed by all votes is indicated in the bottom row, red highlighted totals. 
TPR column gives the true positive rate (recall), while FDR gives the false detection rate (1-precision), which is the percentage of alerts that are false positives. 
Shape sequences were not sufficiently long on Idle task runs to apply Data Smashing. 

\item[2] Analyzing the unsupervised results reveals perfect detection is obtained by the overall classifier, i.e., by using all three tasks' votes.  
\item[3] Delving into each task shows perfect detection by the IE task alone and the Idle task alone, while the Registry task alone exhibits perfect true positive results with only a 0.0625 false detection rate (one false positive of sixteen alerts).  

\item[4] Zooming into the features that contribute least, notice first that the permutation entropy feature never deviates from the training mean by more than a standard deviation. 
Qualitatively this means that the patterns of variation of the profiles over $6\times .01$s$ = .06$s is not changed in a noticeable way by the rootkits tested. 
Quantitatively, this means that inclusion of permutation entropy does not affect the detection outcomes. 

\item[5] Notice that DSD distance, both for shape symbol sequences and on the two-symbol representation of the power data struggle to beat the random baseline.  
Our observation is that this method, while mathematically exciting, is not a good detector for the detection task at hand. 

\item[6] Finally, we discuss the features exhibiting strong classification performance. 
Inspecting the four moments we see that individually on each task all have perfect true-positive rate and generally low false detection rate.   
Similar results are exhibited by the $L^2$-norm feature. 
Hence, the distribution of the power data, regardless of the time progression (encoded by the moments), and the overall shape of the profiles (encoded by $L^2$-norm difference from the baseline profile) are the heavy lifters contributing to the correct classification.  

\end{tablenotes}
\end{threeparttable}

\end{landscape}
\restoregeometry 

\noindent
training profiles of each task.
Our ensemble classifier is constructed from multiple instances of a simple single-feature anomaly detector, all obeying a principled threshold universally\textemdash any observation that is at least one standard deviation
from the mean is considered anomalous, where the standard deviation and the mean are computed on the training observations.
Explicitly, for a given feature, let $V = (v_i: i = 1, ..., n)$ be the training runs' values, and let  $m, s$ be the the the mean and standard deviation of $V$, respectively.
Then given an  observation of this feature, $v$,  from either a training or testing run, we compute the z-score normalization, $z = |v-m|/s$, and vote if $z \geq 1$.
Thus, each of the features informs the single-feature detector, and their tallies are the total votes in favor of alerting on this profile.
The number of votes are in turn the lone feature for the final anomaly detector, which makes the classification  via the same criteria.
The mean and standard deviation of the training runs' vote totals are computed; each
run's vote count is converted to a z-score;
those runs with $z \geq 1$ are labeled infected.
Since computation of permutation
entropy features requires explicit estimation of a probability distribution over the training data permutations, we can easily access $m$, the mean information (permutation entropy of training data), and $s$, the standard deviation of the information over the training runs as follows,  $m:= \sum_{i = 1}^{m!} -\log(P(\gamma_i))P(\gamma_i) = H(P)$ and $s:= [\sum_{i = 1}^{m!}[ \log^2(P(\gamma_i)) P(\gamma_i)] - H(P)^2]^{1/2}.$
For permutation entropy z-score computation, $m$, and $s$, as computed above from the learned distribution are used.

The unsupervised detector classifies a run based on the number of votes across all three tasks; consequently, we are able to see the detection capabilities of each task's vote sum and also each feature individually.
Recall from Sec.~\ref{sec:unsupervised} that each feature value is normalized into a z-score using the mean and variance of the ten clean training runs, and $z\geq 1$, indicating a value of more than one standard deviation from the mean, constitutes a vote for this profile being anomalous.
The vote threshold for final classification is also computed as the mean plus one standard deviation of the vote counts for the training runs.
These vote thresholds are then used to decide the classification of the unseen test profiles.

Table~\ref{tab:unsupervised-testing} presents the results along with each test profiles' set of z-scores and the vote thresholds.
Discussion of the results and efficacy of the features occurs in the table notes.
Notice that the test set class bias is 25/75\% uninfected/infected; hence, assigning labels blindly according to this distribution produces an expected true positive rate of 75\% and false detection rate of 25\%.
Even against this normalization, our results are overwhelmingly positive and give strong empirical evidence that the power profiles are changed sufficiently by malware executions to admit detection.
All together, the ensemble does better than any single detector, which clearly exhibits that this ensemble method reduces the false positives of the constituent detectors without sacrificing the true positive rate.
In particular, using the ensemble overcomes the errant votes from the DSD features as well as the ambient noise in the power data from background processes.
Our recommendation in light of this analysis and
the supervised results is to use an unsupervised ensemble detection method as above with only five features, namely the four moments and the $L^2$ distance.

\subsection{Supervised Approach: Kernel SVMs}
To test a supervised approach, we use hold-one-out validation on the rootkits;
i.e., in each fold training is performed on 12 of 15 uninfected profiles plus 12 infected profiles (3 profiles $\times$ 4 rootkits),
and testing is performed on the 3 profiles from the held-out rootikit combined with the remaining three uninfected profiles.
We test kernel SVMs as the kernels allow versatility and non-linear decision boundaries.
Our implementation used SciKit-Learn~\cite{scikit-learn} to test the following six kernels: linear, radial basis function (RBF) with $\gamma = 0.1, 0.01, 0.001$, and polynomial of degree two and three.
The features for the learner match those of the unsupervised approach, Sec.~\ref{sec:features}.

\begin{table}
  \centering
  \caption{Supervised Learning Results}
    \begin{tabular}{lcccccc}
    \toprule
    \multicolumn{1}{c}{SVM Kernel} & TP & FP & TN & FN & TPR & FDR \\
    \midrule
    Linear              & 15    & 3     & 12    & 0     & 1.00  & 0.17 \\
    RBF $\gamma = 0.001$ & 15    & 3     & 12    & 0     & 1.00  & 0.17 \\
    RBF $\gamma = 0.01$  & 15    & 3     & 12    & 0     & 1.00  & 0.17 \\
    RBF $\gamma = 0.1$   & 14    & 3     & 12    & 1     & 0.93  & 0.18 \\
    Polynomial $d = 3$ & 15    & 3     & 12    & 0     & 1.00  & 0.17 \\
    Polynomial $d = 2$ & 15    & 3     & 12    & 0     & 1.00  & 0.17 \\
    \bottomrule
    \end{tabular}%
  \label{tab:supervised-results}%
\end{table}%

Table~\ref{tab:supervised-results} depicts micro-averaged results, showing perfect or near perfect true positive rate (TPR), with 17-18\% false detection rates (FDR).
As the test sets comprised of non-biased classes (50\% clean/50\% infected) a random classification would expect 50\% TPR and FDR. 
These results give further empirical evidence that the power profiles are changed in a detectable way by malware.
Our conclusion is that the one-class anomaly detection is more suited for this capability.

\section{Conclusion} 
\label{sec:conclusions}
In this work we investigate a power-based malware detection approach for general-purpose computers. 
By monitoring CPU power consumption under a fixed set of tasks, we programmatically gathered power data in the presence and absence of five rootkits.  
To characterize the power data, statistical moments in conjunction with more sophisticated time-series-analytic features are used. 
We test both a one-class anomaly detection ensemble (only training on uninfected data) and many kernel SVMs (trained on both uninfected and infected runs, but tested on data from unseen malware samples and uninfected profiles).  
Our results allow analysis of each feature and task, but the overall result is that perfect detection is achievable (at least in our test environment) by the one-class method. 

Our work proves the concept that rootkit actions change the power profile of the CPU in a detectable manner. 
Because of the high variability of workloads that general-purpose computers undergo, using anomaly detection on ambient power usage is not a viable option (at least from our results and method). 
However, we hypothesize that certain malware can be induced to reveal its presence,  and have proven that baselining the power consumption of a fixed set of tasks is a potential avenue for future detection technology.  
The tasks in our experiment (especially IE), may experience high variability over time in a real-world setting (e.g., added IE plugins, resolving different URLs, etc.) that could create false positives. 
Therefore, our next steps will involve crafting/testing tasks that promise a predictable power curve in a non-lab environment and can induce rootkits action (e.g., printing a small fixed section of the registry).  
We note that accurate results were possible with relatively slow sampling rates.\footnote{Power sampled at $\mathcal{O}$100Hz, implying that even a modest 1GHz processor will exhibit approximately $10^{-9}/10^{-2} = 10$ million cycles between each power data sample.}  
This indicates that the malware tested is not subtle when executing, but our sampling rates may be insufficient for detection of future malware. 
Research to increase the sampling rate is necessary, especially for accurate baselining of very small fixed instructions sequences. 
Overall, we believe this work is a valuable step towards a promising option for malware detection.

\section*{Acknowledgments}


The authors thank Darren Loposser from the Research Instrumentation Group at ORNL, for his contributions to this project, providing electronics and sensors support.

Research sponsored by the Laboratory Directed Research and Development Program of Oak Ridge National Laboratory, managed by UT-Battelle, LLC, for the U. S. Department of Energy. This material is based upon work supported by the U.S. Department of Energy, Office of Energy Efficiency and Renewable Energy, Building Technologies Office.


\bibliographystyle{IEEEtran}
\bibliography{IEEEabrv,refs}

\begin{thebibliography}{10}
\providecommand{\url}[1]{#1}
\csname url@samestyle\endcsname
\providecommand{\newblock}{\relax}
\providecommand{\bibinfo}[2]{#2}
\providecommand{\BIBentrySTDinterwordspacing}{\spaceskip=0pt\relax}
\providecommand{\BIBentryALTinterwordstretchfactor}{4}
\providecommand{\BIBentryALTinterwordspacing}{\spaceskip=\fontdimen2\font plus
\BIBentryALTinterwordstretchfactor\fontdimen3\font minus
  \fontdimen4\font\relax}
\providecommand{\BIBforeignlanguage}[2]{{%
\expandafter\ifx\csname l@#1\endcsname\relax
\typeout{** WARNING: IEEEtran.bst: No hyphenation pattern has been}%
\typeout{** loaded for the language `#1'. Using the pattern for}%
\typeout{** the default language instead.}%
\else
\language=\csname l@#1\endcsname
\fi
#2}}
\providecommand{\BIBdecl}{\relax}
\BIBdecl

\bibitem{9}
{S.S. Clark et al.}, ``Wattsupdoc: Power side channels to nonintrusively
  discover untargeted malware on embedded medical devices,'' in
  \emph{HealthTech}, 2013.

\bibitem{gonzalez2009power}
C.~R.~A. Gonz\'{a}lez and J.~H. Reed, ``Power fingerprinting in {SDR \& CR}
  integrity assessment,'' in \emph{MILCOM}.\hskip 1em plus 0.5em minus
  0.4em\relax IEEE, 2009, pp. 1--7.

\bibitem{gonzalez2011}
C.~R.~A. Gonz{\'a}lez and J.~H. Reed, ``Power fingerprinting in {SDR} integrity
  assessment for security and regulatory compliance,'' \emph{Analog Integrated
  Circuits and Signal Processing}, vol.~69, no. 2-3, pp. 307--327, 2011.

\bibitem{gonzalez2014detecting}
C.~A. Gonz\'{a}lez and A.~Hinton, ``Detecting malicious software execution in
  programmable logic controllers using power fingerprinting,'' in \emph{Inter.
  Conf. on Critical Infrastructure Protection}.\hskip 1em plus 0.5em minus
  0.4em\relax Springer, 2014, pp. 15--27.

\bibitem{reed2012}
J.~H. Reed and C.~R.~A. Gonzalez, ``Enhancing smart grid cyber security using
  power fingerprinting: Integrity assessment and intrusion detection,'' in
  \emph{{FIIW}}.\hskip 1em plus 0.5em minus 0.4em\relax IEEE, 2012, pp. 1--3.

\bibitem{zhong2015side}
{X. Zhong et al.}, ``Side-channels in electric power synchrophasor network data
  traffic,'' in \emph{Proc. 10th Ann. {CISRC}}.\hskip 1em plus 0.5em minus
  0.4em\relax ACM, 2015.

\bibitem{8}
J.~Hoffmann, S.~Neumann, and T.~Holz, ``Mobile malware detection based on
  energy fingerprints - {A} dead end?'' in \emph{Int. Workshop on RAID}.\hskip
  1em plus 0.5em minus 0.4em\relax Springer, 2013, pp. 348--368.

\bibitem{axelsson2000base}
S.~Axelsson, ``The base-rate fallacy and the difficulty of intrusion
  detection,'' \emph{ACM Trans. Inf. Syst. Secur.}, vol.~3, no.~3, pp.
  186--205, Aug. 2000.

\bibitem{bridges2015multi}
R.~Bridges, J.~Collins, E.~Ferragut, J.~Laska, and B.~Sullivan, ``Multi-level
  anomaly detection on time-varying graph data,'' in \emph{Proc.
  {ASONAM}}.\hskip 1em plus 0.5em minus 0.4em\relax ACM, 2015, pp. 579--583.

\bibitem{bridges2016multi}
------, ``A multi-level anomaly detection algorithm for time-varying graph data
  with interactive visualization,'' \emph{Soc. Netw. Anal. \& Mining}, vol.~6,
  no.~1, p.~99, 2016.

\bibitem{christodorescu2007can}
M.~Christodorescu and S.~Rubin, ``Can cooperative intrusion detectors challenge
  the base-rate fallacy?'' in \emph{Malware Detection}.\hskip 1em plus 0.5em
  minus 0.4em\relax Springer, 2007, pp. 193--209.

\bibitem{ferragut2011automatic}
E.~Ferragut, D.~Darmon, C.~Shue, and S.~Kelley, ``Automatic construction of
  anomaly detectors from graphical models,'' in \emph{CICS}.\hskip 1em plus
  0.5em minus 0.4em\relax IEEE, 2011, pp. 9--16.

\bibitem{ferragut2012new}
E.~M. Ferragut, J.~Laska, and R.~A. Bridges, ``A new, principled approach to
  anomaly detection,'' in \emph{{11th ICMLA}}, vol.~2.\hskip 1em plus 0.5em
  minus 0.4em\relax IEEE, 2012, pp. 210--215.

\bibitem{harshaw2016graphprints}
C.~Harshaw, R.~Bridges, M.~Iannacone, J.~Reed, and J.~Goodall, ``Graphprints:
  Towards a graph analytic method for network anomaly detection,'' in
  \emph{Proc. of the 11th CISRC}.\hskip 1em plus 0.5em minus 0.4em\relax New
  York, NY, USA: ACM, 2016, pp. 15:1--15:4.

\bibitem{moore2017modeling}
M.~Moore, R.~Bridges, F.~Combs, M.~Starr, and S.~Prowell, ``Modeling
  inter-signal arrival times for accurate detection of can bus signal injection
  attacks,'' in \emph{Proc. 12th {CISRC}}.\hskip 1em plus 0.5em minus
  0.4em\relax ACM, 2017, pp. 11:1--11:4.

\bibitem{thomas2010rapid}
A.~Thomas, ``Rapid: Reputation based approach for improving intrusion detection
  effectiveness,'' in \emph{{6th IAS})}.\hskip 1em plus 0.5em minus 0.4em\relax
  IEEE, 2010, pp. 118--124.

\bibitem{Hoglund08}
G.~Hoglund and J.~Butler, \emph{Rootkits Subverting the Windows Kernel}.\hskip
  1em plus 0.5em minus 0.4em\relax Addison Wesley, 2008.

\bibitem{koller2010wattapp}
R.~Koller, A.~Verma, and A.~Neogi, ``Wattapp: an application aware power meter
  for shared data centers,'' in \emph{Proc. of the 7th ICAC}.\hskip 1em plus
  0.5em minus 0.4em\relax ACM, 2010, pp. 31--40.

\bibitem{Feng2005}
X.~Feng, R.~Ge, and K.~W. Cameron, ``Power and energy profiling of scientific
  applications on distributed systems,'' in \emph{Proc. of the 19th IEEE Int.
  IPDPS}.\hskip 1em plus 0.5em minus 0.4em\relax Washington, DC, USA: IEEE
  Comp. Society, 2005, p.~34.

\bibitem{Polakis2015PowerslaveAT}
I.~Polakis, M.~Diamantaris, T.~Petsas, F.~Maggi, and S.~Ioannidis,
  ``Powerslave: Analyzing the energy consumption of mobile antivirus
  software,'' in \emph{DIMVA}, 2015.

\bibitem{Azmoodeh2017}
A.~Azmoodeh, A.~Dehghantanha, M.~Conti, and K.-K.~R. Choo, ``Detecting
  crypto-ransomware in {I}o{T} networks based on energy consumption
  footprint,'' \emph{JAIHC}, 2017.

\bibitem{bridges2017setting}
R.~A. Bridges, J.~D. Jamieson, and J.~W. Reed, ``Setting the threshold for high
  throughput detectors: A mathematical approach for ensembles of dynamic,
  heterogeneous, probabilistic anomaly detectors,'' in \emph{2017 IEEE
  International Conference on Big Data (Big Data)}, Dec 2017, pp. 1071--1078.

\bibitem{chattopadhyay2014data}
I.~Chattopadhyay and H.~Lipson, ``Data smashing: Uncovering lurking order in
  data,'' \emph{J. Royal Soc. Interface}, vol.~11, no. 101, p. 20140826, 2014.

\bibitem{cao2004detecting}
{Cao, Y. et al.}, ``Detecting dynamical changes in time series using the
  permutation entropy,'' \emph{Physical Review E}, vol.~70, no.~4, p. 046217,
  2004.

\bibitem{118}
E.~Rodionov and A.~Matrosov, ``The evolution of {TDL}: {C}onquering x64,''
  Retrieved from:
  \url{http://go.eset.com/resources/white-papers/The_Evolution_of_TDL.pdf}.

\bibitem{TracerDAQPro}
MC-MeasurementComputing, ``Tracer{D}{A}{Q}{P}ro,''
  \url{www.mccdaq.com/daq-software/tracerdaq-series.aspx}.

\bibitem{hernandez2017}
\BIBentryALTinterwordspacing
J.~M.~H. Jim{\'{e}}nez, J.~A. Nichols, K.~Goseva{-}Popstojanova, S.~J. Prowell,
  and R.~A. Bridges, ``Malware detection on general-purpose computers using
  power consumption monitoring: {A} proof of concept and case study,''
  \emph{CoRR}, vol. abs/1705.01977, 2017. [Online]. Available:
  \url{http://arxiv.org/abs/1705.01977}
\BIBentrySTDinterwordspacing

\bibitem{ref:alureon1}
Mandiant, ``{T}echnical {R}eport{ M-T}rends 2015,''
  \url{www2.fireeye.com/rs/fireye/images/rpt-m-trends-2015.pdf}.

\bibitem{ref:Pihar1}
Remove-Malware-Tips, ``Get rid of pihar.{B} {T}rojan,'' Retrieved from:
  https://goo.gl/UjJzd7.

\bibitem{ref:Xpaj1}
B.~Miller, ``Got infected with rootkit.mbr.{X}paj? {H}ow to remove it?''
  Retrieved from: \url{https://goo.gl/WKopjC/}.

\bibitem{ref:ZeroAccess1}
Symantec, ``Trojan.{Z}eroaccess,'' Retrieved from: \url{https://goo.gl/FW7xn1}.

\bibitem{ref:MaxRootkit1}
``Rootkit.win32.zaccess.c,'' Retrieved from:
  \url{www.enigmasoftware.com/rootkitwin32zaccessc-removal}.

\bibitem{ref:TDSSkiller}
``{TDSS Killer},'' \url{http://usa.kaspersky.com/downloads/TDSSKiller}, 2016.

\bibitem{ref:WindowsRootkitRemovalTool}
Microsoft, ``{M}alicious {S}oft {R}emoval {T}ool,''
  www.microsoft.com/en-us/download/malicious-software-removal-tool-details.aspx.

\bibitem{ref:VxStreamSandbox}
P.~Security, ``{V}x{S}tream {S}andbox - {A}utomated {M}alware {A}nalysis
  {S}ystem,'' 2016, www.payload-security.com/.

\bibitem{tibshirani2001estimating}
R.~Tibshirani, G.~Walther, and T.~Hastie, ``Estimating the number of clusters
  in a data set via the gap statistic,'' \emph{Journal of the Royal Stat. Soc.:
  Ser.B (Stat. Method.)}, vol.~63, no.~2, pp. 411--423, 2001.

\bibitem{zhang2015novel}
{Zhang X. et al.}, ``A novel bearing fault diagnosis model integrated
  permutation entropy, ensemble empirical mode decomposition and optimized
  {SVM},'' \emph{Measurement}, vol.~69, pp. 164 -- 179, 2015.

\bibitem{wen2012vector}
Y.~Wen and A.~Ray, ``Vector space formulation of probabilistic finite state
  automata,'' \emph{Journal of Computer and System Sciences}, vol.~78, no.~4,
  pp. 1127--1141.

\bibitem{scikit-learn}
{F. Pedregosa et al.}, ``Scikit-learn: Machine learning in {P}ython,''
  \emph{JMLR}, vol.~12, pp. 2825--2830, 2011.

\end{thebibliography}

\end{document}